\documentclass[11pt,twoside]{article}

\usepackage{asp2006}
\usepackage{fancyhdr,graphicx,caption,rotating,multirow,lscape}
\DeclareGraphicsExtensions{.eps,.eps.gz,.ps,.jpg,.tiff}

\begin{document}

\title{The Life Cycle of an XO Planet and the\\
Potential to Detect Transiting Planets of M Dwarfs}

\author{Peter R. McCullough}
\affil{Space Telescope Science Institute, Baltimore, MD 21218, USA}
\author{Christopher J. Burke}
\affil{Space Telescope Science Institute, Baltimore, MD 21218, USA}

\begin{abstract} 
We describe strategies and tactics for detecting transiting planets, as learned from the
experience of the XO Project. A key component is the web-enabled collaboration with a
longitudinally-distributed
Extended Team of dedicated volunteers operating small-aperture telescopes near their homes.
We also quantify the (small) potential to discover
transiting planets of M dwarfs from existing data such as that obtained by the XO Project.
\end{abstract}


\section{Preface}   

Participants in the {\it Workshop} and the reader (we hope) generally
will be aware of the many facets of detection and characterization of
transiting planets.  The scope of this contribution is only to describe
some aspects of our incremental contributions to the art, strategies,
and tactics of the particular topic of detecting transiting planets,
as learned from our XO experience.  Others may or may not find the
tactics described here appropriate for their particular circumstances.
In particular, a team with institutional access to a telescope and
spectrograph and many nights of time allocated to their follow-up
spectroscopy, logically could adopt tactics different from those we have
adopted.  Our aim has been to obtain precision multi-color photometry of
transiting planet candidates to enable us to use efficiently $\sim 20$
hours allocated per year on a large telescope with an excellent
spectrograph. That is, we aim to discriminate as many candidates as
practical with follow-up photometry, prior to spectroscopy.

\section{Life Cycle of an XO Planet}   
\label{sec:lifecycle}

The XO Survey telescope and its operation and data analysis have been
described by McCullough et al. (2005).  In summary, the XO observatory
monitors tens of thousands of bright (V$<12$) stars twice every ten
minutes on clear nights for more than 2 months per season of visibility
for each particular star. XO has been operational since September 2003,
and we observe each star for two seasons before moving on to new fields
of view.  There are two reasons for observing for more than one season:
1) to increase the probability of detection of a transiting planet, and
2) to enable multi-year precision for the extrapolation of ephemerides
for follow-up observations, either precision photometry or spectroscopy.

\begin{figure}[!ht]
\centering
\includegraphics[angle=0,width=15cm]{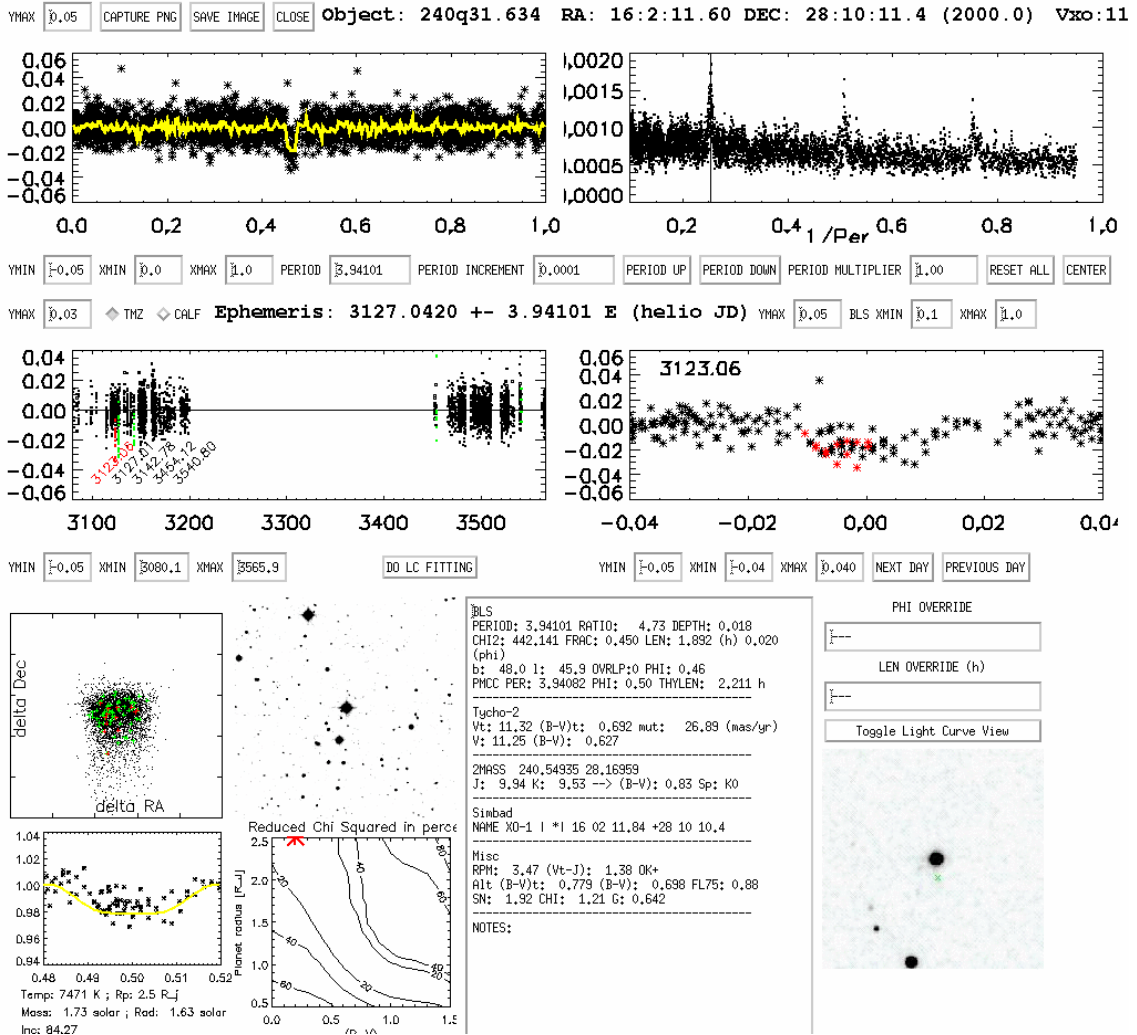}
\caption{The ``Wanted Poster'' for XO-1.\label{fig:xo1poster}}
\end{figure}

From our analysis of more than 3000 observations per star over two
seasons, we identified XO-1 as one of dozens of stars with light curves
suggestive of a transiting planet.  Figure \ref{fig:xo1poster}) is an
example of a ``Wanted Poster'' that software scripts\footnote{Developed
in IDL by Scott Fleming, Jeff Valenti, and Christopher Burke.} create
for each candidate for which we might consider follow-up photometry
or spectroscopy. The example is XO-1, although at the time of its
identification, its ``Wanted Poster'' was much less sophisticated than
the one illustrated in Figure \ref{fig:xo1poster}), which represents
the state of the art as of September, 2006.

Here we describe those elements of the ``Wanted Poster'' that are
not self-evident or that are critical to the method of identifying
excellent candidates from the many others. Each poster is available
to observers as a static image on a password-protected website,
and to its creators as a human-interactive IDL widget.  The upper left
plot (differential magnitude versus phase) is the calibrated XO survey
photometry folded with the period identified by the Box Least Squares
algorithm, (BLS, Kov{\' a}cs et al. 2002), and below it is
the unfolded version of the same photometry (w.r.t. Julian date minus
2450000). The upper right plot is the BLS spectrum with the ``best''
period indicated by the vertical line.  Below that is the folded light
curve again, but zoomed into the time around transit.  In the lower left
quadrant, there are three plots (a centroid plot, a fit to the light
curve\footnote{In this case, the fit to the XO-1 light curve is awful as it does not use the B-V constraint known for the star.}, and a plot of $\Delta \chi^2$) and a finder chart centered on the
target star from the Digital Sky Survey, DSS.  The centroid plot is a
scatter diagram of all centroids measured from the XO survey data. Colored
points\footnote{The proceedings may not be in color.} are those obtained
at times predicted to be during transit. If the colored points appear
significantly offset from the swarm of uncolored points, i.e. a centroid
shift is evident during the transit, we suspect the candidate to be a
blend of a bright star with a fainter eclipsing binary.  The direction
of the centroid shift vector predicts the direction of the blending star,
and the vector's magnitude approximately predicts
its separation.\footnote{The former 
may be intuitive but the latter counterintuitive, given the general
condition that the blending eclipsing binary star's characteristics are
unknown. The key is that the depth of the dip in the combined light is small and
known.} Only rarely will such a blend be a case of a transiting planet:
an example might have been HAT-P-1 (a double star of two nearly-identical
stars separated by 11\arcsec, with one star transited by a hot Jupiter;
Bakos et al. 2006) observed with survey images of angular resolution
of 14\arcsec.  To assist in discriminating the two cases, we reproduce a
finder chart from 2MASS (lower right of ``Wanted Poster''), because it has
better angular resolution than the DSS, and from the 2MASS point-source
catalog we estimate the faction of light (FL75 on the ``Wanted Poster'')
contained by the single-brightest-star within XO's photometry aperture
of 75\arcsec\ radius.  From the FL75, we can reconstruct light curves
to determine if a candidate is plausibly a transiting planet. Although
such light curves can be studied interactively, we have also found it
helpful to contour $\Delta \chi^2$ of the observations w.r.t. potential
models, parameterized by the radius of the hypothetical planet (in R$_J$)
and the B-V color of a main-sequence star; one such plot is below the DSS
finder chart on the ``Wanted Poster.'' Then at a glance, a person can
take the observed B-V color and its uncertainty (evident on the poster
by B-V's various forms inferred from Tycho-2, 2MASS, etc) and determine
the likely range of planet radii required to match the survey photometry.
Other useful facts provided on the poster are the Galactic longitude and
latitude (there are more astrophysical false positives at low galactic
latitudes), the length of the transit as reported by BLS (LEN) or in
theory for a central-transit of a sun-like star (THYLEN), the proper
motion and the Reduced Proper Motion \citep{GOU03} criterion (OK$+$ indicating to us
that the candidate star is small enough to be consistent with a $\sim 1$\%
transit depth from a hot-Jupiter-sized planet).

\begin{figure}[!ht]
\centering
\includegraphics[angle=0,width=12cm]{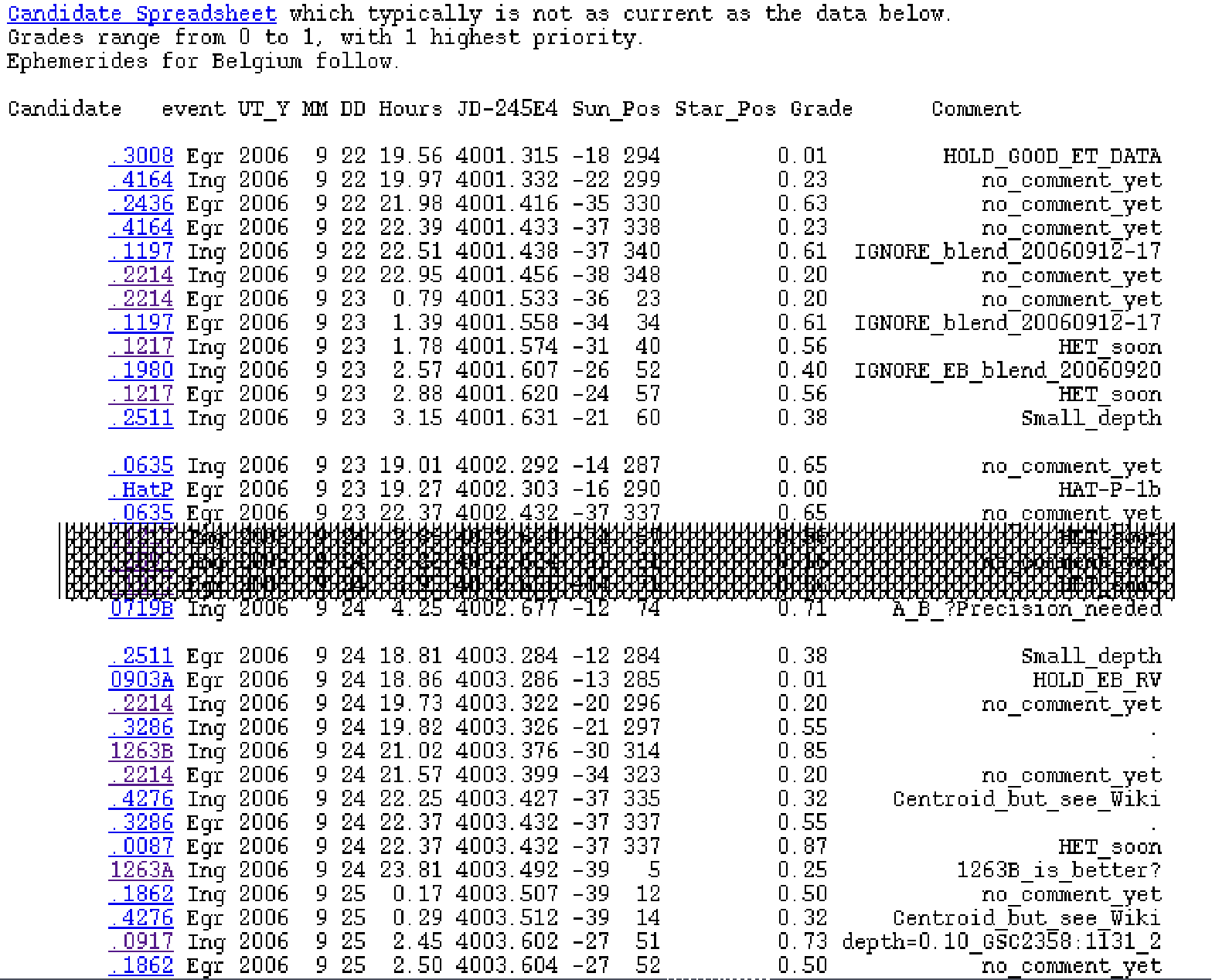}
\caption{Ephemerides provided to the XO Extended Team. Candidate names,
and positions (altitudes and azimuths) have been redacted.
The ``event'' column lists Egr(esses) and Ing(resses).
On the web, target names are hyperlinked to the corresponding
``Wanted Poster'' (cf. Figure \protect{\ref{fig:xo1poster}}).
\label{fig:ETwebsite}}
\end{figure}

Every few months, we analyzed and winnow many dozen ``Wanted Posters''
to a few dozen that we post on a password-protected website accessible
by the XO Extended Team (ET). We generate custom ephemerides for
each ET observer of all active candidates; Figure \ref{fig:ETwebsite}
is an example of such for the nights of Sep 22-25, 2006.  
Each candidate is assigned a grade, which represents a suggested priority
for ET observations.  Grades are
generated initially using a sort of ``Drake Equation'' for a (highly
nonlinear) function
that is related to the probability that a candidate is a transiting
planet. The Drake Equation analogy is apt, because the grade is
(approximately) the product of many probabilities that are often not
more than guesses, automatically generated from the facts on the ``Wanted
Poster.'' 
We manually adjust grades as additional information becomes
available, such as follow-up photometry or spectroscopy. 
Updates to the ephemerides, grades, and commentary occur as required,
typically weekly.

The feedback from the ET typically is both informative and prompt. It is
not uncommon for a new candidate to be rejected or increased markedly
in priority within days of its initial posting to the ET. Prior
to July 2006, the XO Project collaborated with the ET primarily
via a website that provided data to the ET, members of which would
communicate their observations via email to the Principal Investigator,
who added them to the website. A significant improvement to the XO-ET
collaboration was upgrading from a one-to-many website (predicated on
the ET-to-PI email) that was an undesirable throttle on progress,
to a Wiki website that permitted many-to-many collaboration, since the
longitudinally-distributed ET collectively never stops (in principal if
not quite in practice) observing, analyzing data, and collaborating.
The password-protected Wiki allows ET members to upload their observations
in the form of plots of light curves and associated tables of data.

The usual criteria for discriminating candidates from follow-up observations
have been described by others and will not be repeated here. We can
describe a few refinements that we have used and mention a few additional
astrophysical false positives that we have considered, including
triple and quadruple systems, potentially with eccentric orbits.

ET observations have proved very helpful in identifying
gravitationally-bound triple stars, In particular, the few candidates
with XO survey light curves that have photometrically-determined periods
less than 24 hours have all been rejected, to date, based upon multi-color
transit light curves, expertly observed in a timely fashion by the ET. In
some cases, we have verified the triple-star nature spectroscopically,
via small-depth satellite spectral lines near the main (deeper) lines,
or at least have verified no discernible radial velocity variation of the
spectral lines, consistent with the triple-star hypothesis or a very low
mass (and hence an unrealistically low density) planet.  The spectroscopic
signature requires a large telescope and sometimes can be difficult to discern.
The signature of a triple star in a multi-color transit light curve can
be observed with a small telescope and can be obvious:
the ``transit'' depth might be 2\% in R band, 1.5\% in V band, and 1\%
in B band, for example. In some cases, multiple-star systems produce
achromatic ``transit'' depths and shapes, and those systems require
spectroscopy.

For planets with periods longer than $\sim$1 week, the orbit may not
have been circularized by tidal effects. Implications of eccentricity
will be analyzed elsewhere (Burke 2007, in preparation).  Here we note one additional
astrophysical false positive that we have imagined: an eclipsing binary
system of two stars of identical effective temperature, in which the
smaller star passes behind the larger star, causing the combined light to
dip, but on the near-side passage, the smaller star misses transiting the
larger star due to the orbital eccentricity. Such a system, improperly
interpreted, can produce a {\it negative mass} for the ``transiting
planet'' because the radial velocity curve is 180\deg\ out of phase of
that expected by the transit model.

Experience has forced us to acknowledge that there will be some planet
candidates for which measurement of the planet's mass is impossible
spectroscopically, not simply because the host star is too faint (e.g Sahu
et al. 2006), but even for bright stars such as the XO Project observes,
because the spectral lines of the host star are very
rotationally-broadened, e.g. an F dwarf with Vsini $> 30$km s$^{-1}$.
An upper limit might be achievable that ``proves'' the object's mass
is substellar, and the Rossiter effect could discriminate between 1) a
triple star and 2) a transiting planet-sized object. However, the mass
of the latter object could be unknown and (with current techniques)
unknowable and yet the upper limit might not discriminate between a
brown dwarf and a planet.

A metamorphosis occurs for the rare
and nearly unknown star (an egg) to that of a star
of intense scrutiny by many astronomers (the butterfly).
Here we have described only the caterpillar phase of the life cycle.
It changes markedly with two or three 
precision radial velocities exhibiting sinusoidal variation of
$\sim$0.1 km s$^{-1}$ \citep{MCC06}.
At that point it enters the pupa phase, involving preparation and planning for
flight operations, such as on HST and SST, and for the day when its discoverers
reveal it to all as a beautiful gift of Nature.

\section{Potential to Detect Transiting Planets of M Dwarfs}  

Finding planets around M dwarfs has several scientifically rewarding
benefits.  Foremost, the small stellar radii and masses of M dwarfs
provide sensitivity to smaller-radius planets in the case of transit
searches and to lower-mass planets in the case of radial velocity
searches.  Also, as emphasized in the recent review by \citet{TAR06}, a
full consensus of life in the Universe would not be complete without
planet statistics for M dwarfs that compromise 75\% of all stars in
the Galaxy.  Finally, the habitable zone for M dwarfs is at small
separations (0.02-0.2 AU) and periods (days, not years) improving
detectability.  Thus, searches for planets of M dwarfs offer
great potential to study and characterize potential life-bearing
planets.

In addition to the scientific benefits of M dwarf planet searches, M
dwarfs offer a technical benefit to transiting planet surveys.  As
opposed to FGK spectral types, the M dwarfs are readily separated from
giant contaminants that plague transit surveys.  The radii of
giants across all spectral types are too large to find transiting
planets with $\sim$1\% photometry typical for current, ground-based
surveys.  For a V$<$11 transit survey, only $\sim$10\% of stars are dwarfs
\citep{GOU03}.  Thus, separating dwarf from giant
stars is imperative.  In particular, photometric color-color diagrams
can separate dwarfs from giants in the M dwarf regime but not well
for FGK spectral types.

Another method of separation makes use of the $\sim$100x luminosity ratio
between M giants and M dwarfs (5 magnitudes difference). For an M dwarf and an M giant of similar
color and apparent magnitude, the dwarf must be much closer and consequently
will have a higher
proper motion than the giant assuming similar space velocity.  
The so-called Reduced Proper Motion diagram \citep{GOU03}
is a powerful method of dwarf/giant discrimination for stars
with cataloged proper motions and photometric colors, and it
is particularly effective for M spectral types.
The smaller
brightness contrast between dwarf and giant results in a less
effective separation in the reduced proper motion diagram for earlier
(FGK) spectral types.

Although currently limited to the North Celestial Sphere, the
\citet{LEP05} proper motion catalog (LSPM) provides an excellent
source of M dwarf candidates.  The catalog has a limiting proper
motion of $>$150 mas yr$^{-1}$ with is more than 90\% complete for
$|b|>15$\deg\ galactic latitude down to V$<$19 mag.  In addition,
the catalog has been matched to the 2MASS catalog and each
entry already has been verified individually by a human.

The star counts in the LSPM catalog provide an excellent means to
generally characterize the properties of a transit survey necessary to
detect a robust population of transiting planets around M dwarfs.
Assuming the fraction of M dwarfs with a Hot Jupiter (P$<3$
day) planet, $f\sim 0.5$\%, and the probability for a Hot Jupiter to
transit a M dwarf $\sim 9$\%, imply approximately 1 transiting Hot
Jupiter for every 2200 M dwarfs observed.  The fraction of M dwarfs
with Hot Jupiters is uncertain but tentatively 
is less than the fraction of FGK stars with Hot Jupiters ($\sim
1.2$\%) \citep{MAR05,BUT04,END06}.  By selecting M dwarfs in the (J-H)
vs. (H-K) color-color diagram from LSPM catalog, we find only 4800 M
dwarfs with V$<$14, 8700 M dwarfs with V$<$15, and 15000 M
dwarfs with V$<$16.  This implies at least 2, 4, and 7 transiting Hot
Jupiters orbit M dwarfs with V$<$14, V$<$15, and V$<$16,
respectively in the North Celestial Sphere.

An additional caveat is relevant to the LSPM
catalog.  The lower limit to the proper motions in the catalog 
increasingly rejects luminous M dwarfs at large distances,
i.e. fainter
apparent magnitudes. For example, a typical local motion of 40 km s$^{-1}$ corresponds
to a maximum distance of 50 pc for the proper motion to be greater than 150 mas
yr$^{-1}$, the LSPM catalog's lower limit. An M0 dwarf with $M_{V}=9$ has $V=12$
at a distance of 50 pc, hence for magnitudes V$>12$, the catalog will begin to be increasingly incomplete for M0 dwarfs toward fainter magnitudes (i.e. an M0 dwarf is
further than 50 pc and its proper motion will be less than the
catalog's limit). 

Despite the advantage the faintness of M dwarfs provide in making them
readily detected in proper motion catalogs, their faintness
drastically hinders finding transiting planets around M dwarfs in a
bright star transit survey such as XO \citep{MCC05}.  XO transit
survey is a bright star transit survey which was designed to find
transiting planets around FGK stars between 9$<V<$12.  The M dwarf
statistics from the LSPM catalog show that the XO survey is not
sensitive enough to discover more than a very few
transiting planets of M dwarfs, unless the fraction of hot Jupiters
($f\sim 0.5$\%)
around such stars has been significantly underestimated by the radial
velocity surveys to date. Surveys such as XO
might effectively identify {\it transits}
in the case of a planet with period and
phase known {\it a priori} from a radial velocity survey,
because those parameters limit the search space and consequently improve the
detection sensitivity, so an M dwarf's faintness isn't as limiting in that
case.
Problems related to blending of light from additional star(s) described
in Section 2 are particularly acute for (faint) M dwarfs.

A single XO strip covers 440 sq. deg. and on average contains 40 M
dwarfs from the LSPM catalog down to V$<$12.8 limit of XO.  After
calculating the blending in the photometric aperture for M dwarfs in
an XO field of view only 20 M dwarfs are unblended ($>$75\% of the light
in the aperture arises from the the M dwarf).  Overall, XO currently
has data available for $\sim$ 200 M dwarfs, and after complete
coverage of the North Celestial Sphere $\sim$ 800 M dwarfs.  Although
not routinely analyzed for transiting planets around FGK stars, XO
data down to V$<$14 have sufficient accuracy to detect Jupiter radii
planets around M dwarfs.  We estimate further analysis of this deeper
magnitude limit may provide a total of $\sim 1600$ M dwarfs observed
by XO in the North Celestial Sphere.

From this study we conclude a transiting Hot Jupiter around a bright
(V$<$14) M dwarf will be rare. A statistically significant sample of
transiting M dwarf planets will require photometry of fainter stars (V$>$16) and
angular resolving power than existing XO data can provide.
A proper motion catalog complete to smaller
proper motions than LSPM, while still maintaining the laborious human
verification of proper motion and matching to 2MASS would be helpful.

\acknowledgements 
The XO Extended Team is critical to the XO Project's ability to discover
transiting hot Jupiters.
As of September, 2006, contributing members of the XO Extended Team are
Ron Bissinger,
Mike Fleenor,
Cindy Foote,
Enrique Garcia-Melendo,
Bruce Gary,
Paul Howell,
Franco Mallia,
Gianluca Masi,
and
Tonny Vanmunster.
Colleagues currently contributing to the XO Project, but not to any
errors or mis-statements in this manuscript (!), are Ken Janes, Chris
Johns-Krull, Jim Heasley, and Jeff Valenti.

Emeritus contributors to the XO Project are
(in reverse chronological order)
Jeff Stys (data analysis, etc),
Scott Fleming (data analysis, etc),
Knute Ray (hardware technician),
Beth Bye (initial deployment),
and
Chris Dodd (initial deployment).

We thank Frank Summers for access to a Beowulf cluster, and the IT Services
Division of STScI for setting up XO's Wiki website.
The University of Hawaii staff have made the operation on Maui possible; we
thank especially Bill Giebink, Les Hieda, Jake Kamibayashi,
Jeff Kuhn, Haosheng Lin, Mike Maberry, Daniel O'Gara,
Joey Perreira, Kaila Rhoden, and the director of the IFA, Rolf-Peter Kudritzki.
We thank the support staff of the McDonald Observatory for providing
excellent observational capabilities for follow-up spectroscopy.

XO is funded primarily by a NASA Origins grant and in the past has
received financial support from NASA Origins, the Sloan Foundation, the
Research Corporation, the Director's Discretionary Fund of the STScI,
and the US National Science Foundation.


\begin{thebibliography}{}


\bibitem[Bakos et al.(2006)]{2006astro.ph..9369B} Bakos, G.~A., et al.\ 
2006, ArXiv Astrophysics e-prints, arXiv:astro-ph/0609369 

\bibitem[Butler et al.(2004)]{BUT04} Butler, R. P., Vogt, S. S., Marcy, G. W., Fischer, D. A., Wright, J. T., Henry, G. W., Laughlin, G., \& Lissauer, J. J. 2004, \apj, 617, 580

\bibitem[Endl et al.(2006)]{END06} Endl, M., Cochran, W. D., K\''{u}rster, M., Paulson, D. B., Wittenmyer, R. A., MacQueen, P. J., \& Tull, R. G. 2006, \apj, 649, 436

\bibitem[Gould \& Morgan(2003)]{GOU03} Gould, A. \& Morgan, C. W. 2003, \apj , 585, 1056

\bibitem[Kov{\' a}cs et al.(2002)]{2002A&A...391..369K} Kov{\' a}cs, G.,
Zucker, S., \& Mazeh, T.\ 2002, \aap, 391, 369

\bibitem[L\'{e}pine \& Shara(2005)]{LEP05} L\'{e}pine, S. \& Shara, M. M. 2005, \aj, 129, 1483

\bibitem[Marcy et al.(2005)]{MAR05} Marcy, G., Butler, R. P., Fischer, D., Vogt, S., Wright, J. T., Tinney, C. G., \& Jones H. R. A. 2005, Progress of Theoretical Physics Supplement, 158, 24

\bibitem[McCullough et al.(2005)]{MCC05} McCullough, P.~R., 
Stys, J.~E., Valenti, J.~A., Fleming, S.~W., Janes, K.~A., \& Heasley, 
J.~N.\ 2005, \pasp, 117, 783 

\bibitem[McCullough et al.(2006)]{MCC06} McCullough, P. R., et al. 2006, \apj, 648, 1228

\bibitem[Sahu et al.(2006)]{2006Natur.443..534S} Sahu, K.~C., et al.\ 2006, 
\nat, 443, 534 

\bibitem[Tarter et al.(2006)]{TAR06} Tarter, J. C., et al. 2006, preprint (astro-ph/0609799)

\end{thebibliography}
\end{document}